\documentstyle{ichep}
\begin{document}

\title{Lattice Results for Heavy Light Matrix Elements}

\author{A. Soni}
\affil{Brookhaven National Laboratory \\
Physics Department \\
Upton, NY\ \ 11973}

\abstract{Lattice results for heavy light matrix elements are reviewed
and some of their implications are very briefly discussed
\cite{glasgow_talk}. Despite the
fact that in most cases the lattice results for weak matrix elements at
the moment have only a modest accuracy of about 20--30\% they already
have important phenomenological repercussions; e.g.\ for
$V_{td}/V_{ts}$, $x_s/x_d$ and $B\to K^\ast\gamma$.}

\twocolumn[\maketitle]

I. As samples of ``new'' projects, I briefly discuss four projects
\medskip

a. The NRQCD group \cite{nrqcd} reports a preliminary result on the
mass splitting $m_{B^\ast_S} - m_{B_S}$. They use 100 configurations on a
$16^3\times 32$ lattice of the HEMCGC group (dynamical, staggered
$n_f=2$, $am=.01$ and $\beta=5.6$). Their very preliminary result is

\begin{eqnarray}
m_{B^\ast_S}-m_{B_S} & = & [50\pm5({\rm statistical}) \pm10({\rm am})
\nonumber \\
& & \quad \pm 15 ({\rm scale})] MeV \label{mbs}
\end{eqnarray}

\noindent Since rough phenomenological estimates for the splitting
invariable yield about 50 MeV, such a lattice calculation will be very
interesting if the errors can be brought down to 0 (5 MeV).
\medskip

b. The LANL collaboration \cite{lanl} is acquiring many interesting
results via their large scale simulations on the CM-5. So far they have
analyzed 58 configurations of size $32^3\times 64$ at $\beta=6.0$. This
lattice has the distinction of having the largest physical volume used so far
for weak matrix element calculations. Comparison with results previously
obtained on smaller lattices will therefore be very valuable. Here are
some of their preliminary findings:

\begin{eqnarray}
(1) & \quad f_K/f\pi & =  1.163\pm.023; \nonumber\\
(2) & \quad \frac{f^{D\to\pi e\nu}_+ (q^2=0)}{f^{D\to Ke\nu}_+ (q^2=0)}  & =
0.86\pm.08; \label{fkfpi} \\
(3) &  \quad \frac{A^{D\to K^\ast e\nu}_2 (0)}{A^{D\to K^\ast e\nu}_1
(0)} & =  .77\pm .22. \nonumber
\end{eqnarray}

c. The Fermilab Static Group (FSG) \cite{fermi} is now finishing a
comprehensive study with the use of their static method.
They reported interesting results
on $f_B$, $f_{B_S}/f_{B_u}$ and $m_{B_S} - m_{B_u}$. With their method
they find a significant lattice spacing ($a$) dependence for $f_B$
whereas the other two quantities appear insensitive to $a$. Their key
results are:

\begin{equation}
f_B  =  188\pm23({\rm stat}) \pm15 ({\rm sys})^{+27}_{-0} ({\rm extr})
\pm 14(a) \label{fb} \\
\end{equation}
\begin{equation}
f_{B_S}/f_{B_u}  =   1.216 \pm .041\pm.016 \label{fbs}
\end{equation}
\begin{equation}
m_{B_S}-m_{B_u}  =  86\pm12\pm7 ({\rm MeV}) \label{mbstwo}
\end{equation}

d. MILC collaboration \cite{milc} is undertaking a dedicated study of
$f_B$ using primarily the Intel Paragon machine. They have analyzed so
far 40 configurations  ($24^2\times80$, $\beta=6.3$). Their preliminary
results for the decay constants are given below:

\begin{eqnarray}
f_B & = & 174\pm 7 \mbox{ MeV} \nonumber \\
f_{B_S} & = & 198\pm 7 \mbox{ MeV} \nonumber \\
f_D & = & 205\pm 5 \mbox{ MeV} \label{milcons} \\
f_{D_S} & = & 228\pm 5 \mbox{ MeV} \nonumber
\end{eqnarray}

II. A illustrative sample of weak matrix elements results are compiled
in Table~I. The ones mentioned above that are now reporting final
results are included where relevant.

\medskip

III. Phenomenological implications. While in many cases the accuracy of
the existing lattice results is not that great, being typically about
20\%, (an exception is $B_K$ \cite{gupta}) yet they have important
implications already. This is in great part due to the fact in many
cases little reliable information exists about several of these matrix
elements. Thus an answer with even a modest accuracy of 20\% can have
crucial impact. Of course lattice methods will continue to provide
refined results in many cases.

I discuss some of the key implications that emerge. The lattice results
for $f_B$ and the $B$ parameter along with the experimental result on
$B$-$\bar B$ mixing leads to:

\begin{equation}
\left| \frac{V_{td}}{V_{ts}} \right| = .22\pm .08
\end{equation}

\noindent Using the above along with the lattice results
for (see Table~1):

\begin{equation}
f_{B_S}/f_B = 1.16\pm .10 \label{fbstwo}
\end{equation}

\noindent leads to an indication for the expectation for $B_S$-$\bar
B_S$ mixing:

\begin{equation}
\frac{x_s}{x_d} = 18\pm 14. \label{xs}
\end{equation}

Finally, the lattice result for $B\to K^\ast+\gamma$:

\begin{equation}
H^{\rm lattice}_{K^\ast}  \equiv \frac{BR(B\to K^\ast\gamma)^{\rm
lattice}}{BR(b\to s\gamma)} = 6.0\pm1.2\pm3.4\% \label{hlat}
\end{equation}

\noindent has important repercussions. Recall two recent experimental
results \cite{ammar,barish}:

\begin{equation}
BR (B\to K^\ast\gamma) =  (4.5\pm1.5\pm.9) \times 10^{-5} \label{brb}
\end{equation}
\begin{equation}
BR(b\to s\gamma) =  (2.32\pm.51\pm.29\pm.32) \times 10^{-4} \label{brbtwo}
\end{equation}

\noindent Together they imply:

\begin{equation}
H^{\rm expt}_{K^\ast} \simeq (19.4\pm 7.8) \% \label{hexp}
\end{equation}

While the errors in the experiment as well as in the lattice
calculations [8--10] are too large at present to allow one to draw
strong conclusions, the difference between the two is a quantitative
measure of the long-distance contributions. The point is that while the
lattice calculation is by construction of a short-distance piece only,
experiments may also be seeing some long-distance contamination coming,
for example, from $B\to\psi_{\rm virtual} + K^\ast$, $\psi_{\rm
virtual}\to \gamma$. Improved lattice calculations as well as
experiments are therefore highly desirable to quantify the extent of the
long-distance contributions to such important radiative decays.

\bigskip

\leftline{\bf Acknowledgements}
\medskip

I am thankful to Christine Davies for inviting me to give this
mini-review. I have also benefitted from discussions with her and with
Claude Bernard, Sara Collins, Estia Eichten, Rajan Gupta, Paul
MacKenzie, Junko Shigemitsu, Jim Simone, and John Sloan for numerous
discussions. This research was supported in part under the  DOE grant
number DE-AC0276CH00016.

\begin{table*}
\Table{lcp{3.8in}}{
\multicolumn{3}{c}{SAMPLE OF RESULTS FOR HADRON MATRIX} \\
\multicolumn{3}{c}{ELEMENTS FROM QUENCHED LATTICE QCD} \\ \\
QUANTITY & VALUE & \hskip1in AUTHORS (REMARKS) \\ \\
\multicolumn{2}{l}{\underbar{The ``$B$'' Parameters}} \\
$\hat B_K$ & $.825\pm.027\pm.023$ & \hskip1in  Gupta,~Kilcup,~Sharpe
     (Staggered) \cite{gupta}  \\
& $.85\pm.20$ &  \hskip1in ELC  (Wilson) \cite{gavelaone}  \\
& $.86\pm.11\pm.05$ &  \hskip1in Bernard,~Soni (Wilson) \cite{bersoni}  \\
\hline
$\hat B_K$ & $.82\pm.10$ &  \hskip1in Most likely${}\equiv90\%$ CL
     \cite{errors} \\
& & \hskip1in  (inc.\ statistical and systematic errors)\\
\hline
$\hat B_B$ & $1.3\pm.2$ &  \hskip1in Bernard, {\it et al.} \cite{bernardtwo} \\
& $1.16\pm.07$ &  \hskip1in ELC  \cite{gavelatwo} \\
\hline
$\hat B_B\simeq \hat B_{B_S}$ & $1.2\pm.2$ & \hskip1in  Most Likely
     (90\% CL) \cite{errors} \\
\hline \\
\multicolumn{2}{l}{\underbar{The Decay Constants}} \\
$f_K/f_\pi$ & $1.08\pm.03\pm.08$ & \hskip1in  Bernard,~Labrenz,~Soni
     \cite{bernardthree} \\
\hline
$f_D$ (MeV) & $174\pm26\pm46$ &  \hskip1in Bernard, {\it et al}.
     \cite{bernardtwo} \\
& $190\pm33$ &  \hskip1in Degrand,~Loft \cite{degrand}  \\
& $210\pm40$ &  \hskip1in ELC  \cite{gavelatwo} \\
& $185^{+4+42}_{-3-7}$ &  \hskip1in UKQCD  \cite{baxter} \\
& $208\pm9\pm32$ &  \hskip1in Bernard,~Labrenz,~Soni \cite{bernardthree}\\
\hline
$f_D$ (MeV) & $197\pm25$ &  \hskip1in Most Likely (90\% CL) \cite{errors} \\
\hline
$f_{D_S}$ (MeV) & $222\pm16$ & \hskip1in  Degrand,~Loft  \cite{degrand}\\
& $234\pm46\pm55$ &  \hskip1in Bernard, {\it et al}. \cite{bernardtwo} \\
& $230\pm50$ &  \hskip1in ELC \cite{gavelatwo}\\
& $212\pm4^{+46}_{-7}$ &  \hskip1in UKQCD  \cite{baxter} \\
& $230\pm 7 \pm 35$ &  \hskip1in Bernard,~Labrenz,~Soni \cite{bernardthree} \\
\hline
$f_{D_S}$ (MeV) & $221\pm30$ & \hskip1in  Most Likely (90\% CL) \cite{errors}
\\
\hline
$f_B$ (MeV) & $205\pm40$ &  \hskip1in ELC  \cite{abadatwo}\\
& $160\pm6^{+53}_{-19}$ &  \hskip1in UKQCD  \cite{baxter} \\
& $187\pm10\pm 37$ &  \hskip1in Bernard,~Labrenz,~Soni \cite{bernardthree} \\
& $188\pm23\pm15^{+27}_{-0}\pm14$ & \hskip1in FSG \cite{fermi} \\
\hline
$f_B$ (MeV) & $173\pm40$ &  \hskip1in Most Likely (90\% CL) \cite{errors} \\
\hline
$f_{B_S}$ (MeV) & $194^{+6+62}_{-5-9}$ & \hskip1in  UKQCD \cite{baxter} \\
& $207\pm9\pm40$ &  \hskip1in Bernard, Labrenz, Soni \cite{bernardthree}\\
\hline
$f_{B_S}$ (MeV) & $201\pm40$ & \hskip1in  Most Likely (90\% CL) \cite{errors}
\\
\hline
$f_{B_S}/f_B$ & $1.22^{+.04}_{-.03}$ & \hskip1in UKQCD \cite{baxter} \\
& $1.11\pm.02\pm.05$ & \hskip1in Bernard, Labrenz, Soni \cite{bernardthree}\\
& $1.22\pm.04\pm.02$ & \hskip1in FSG \cite{fermi} \\
\hline
$f_{B_S}/f_B$ & $1.16\pm.10$ & \hskip1in Most Likely (90\% CL) \cite{errors} \\
\hline \\
\multicolumn{3}{l}{\underbar{Radiative $B$ Decays}} \\
\multicolumn{2}{l}{$R_{K^\ast} \equiv \frac{\Gamma(B
\to\gamma K^\ast)}{\Gamma(b \to \gamma_s)} = 6.0\pm1.2\pm3.4\%$} &
\hskip1in Bernard,~Hsieh,~Soni \cite{bernard} \\
\multicolumn{2}{l}{\hspace*{1.14in} $8.8^{+28}_{-25} \pm3.0\pm1.0$} & \hskip1in
UKQCD  \cite{bowler} \\
}
\end{table*}

\end{document}